\documentclass[twocolumn,prl]{revtex4}
\usepackage{graphicx}
\begin{document}
\title{Highly entangled quantum systems in $3+1$ dimensions}
\author{Brian Swingle}
\email{bswingle@mit.edu}
\affiliation{Department of Physics, Massachusetts Institute of Technology, Cambridge, MA 02139}
\begin{abstract}
Many systems exhibit boundary law scaling for entanglement entropy in more than one spatial dimension.  Here I describe three systems in $3+1$ dimensions that violate the boundary law for entanglement entropy.  The first is free Weyl fermions in a magnetic field, the second is a holographic strong coupling generalization of the Weyl fermion system, and the third is a strong topological insulator in the presence of dislocations.  These systems are unified by the presence of a low energy description that includes many gapless $1+1$ dimensional modes.  I conclude with some comments on the search for highly entangled states of quantum matter and some potential experimental signatures.
\end{abstract}
\maketitle

\textit{Introduction}.---The structure of entanglement in many body systems provides a useful characterization of these systems.  It underlies the functioning of tensor network approaches to many body systems \cite{mera,peps,terg}.  It is also related to experimentally accessible quantities like current fluctuations \cite{emeter}.  Entanglement entropy, defined as the von Neumann entropy of the reduced density matrix of a chosen subsystem, provides a convenient characterization of entanglement.  The emerging picture is that most many body systems live in a small subset of Hilbert space characterized by a geometric boundary law for entanglement entropy.  Holographic duality has also provided insight into the deeply geometrical nature of entanglement \cite{maldecena,witten,polyakov,holo_ee,holo_ee_f}.  I am interested in highly entangled states of quantum matter that violate the typical boundary law for the entanglement entropy.

Many systems, gapless or not, satisfy a boundary law for entanglement entropy in higher dimensions \cite{arealaw1}.  This boundary law means that the entanglement entropy of a region of linear size $L$ in $d$ spatial dimensions scales as $L^{d-1}$.  There are exceptions.  Systems described by conformal field theory in one spatial dimension are known to violate the boundary in a universal way \cite{geo_ent,eeqft}.  The simplest exception in more than one dimension is provided by free fermions which violate the boundary with an extra logarithmic correction \cite{fermion1,fermion2,fermion3,fermion4,fermion5}.  I have argued that Fermi liquids also violate the boundary law in a universal way \cite{bgs_ferm2}.  This argument is based on earlier intuition ascribing the boundary law violation of free fermions to the gapless one dimensional nature of excitations on the Fermi surface\cite{bgs_ferm1}.

Here I use similar intuition to argue for boundary law violations in a variety of systems in higher dimensions.  I will compute the entanglement entropy for three systems: a Weyl fermion in a magnetic field, a holographic system of many interacting Weyl fermions in a magnetic field \cite{mag_branes}, and some strong topological insulators with finite dislocation density \cite{ti1,ti3,ti4,ti5,ti6,ti7,ti10}.  These systems all share a common feature which unifies the discussion, namely the appearance of a large number of gapless one dimensional modes at low energies.  It is these one dimensional modes which are responsible for the violations of the boundary law.

This paper is organized as follows.  I begin by computing the entanglement entropy of a single free Weyl fermion in a background magnetic field.  Next, I embed this Weyl fermion into a strongly coupled quantum field theory with a holographic dual and compute the entanglement entropy for that system.  I also describe a related setup in strong topological insulators.  I conclude with a discussion of these results in the context of searches for highly entangled states of quantum matter.

\textit{Weyl fermion in a magnetic field}.---Consider a single Weyl fermion $\psi$ charged under a gauge field $A$ with charge $q$ in $3+1$ dimensions.  The equation of motion for this fermion is $\gamma^\mu D_\mu \psi = 0$ with $\gamma^5 \psi = -\psi$ where $D_\mu = \partial_\mu - i q A_\mu$ is the covariant derivative.  Let there be a finite magnetic field, say in the $z$ direction: $F_{1 2} = \partial_1 A_2 - \partial_2 A_1 =B$.  The magnetic field defines a length scale called the magnetic length $\ell_B^2 = 1/B$ (the units are made up by the flux quantum).  On length scales much less than $\ell_B$ the theory looks like a $3+1$ dimensional conformal field theory.  On length scales much bigger than $\ell_B$ the theory becomes effectively $1+1$ dimensional.  Indeed, the Weyl fermion is special because it possesses zero modes that avoid being gapped by the magnetic field.

I take the gamma matrices to satisfy $\{ \gamma^\mu , \gamma^\nu \} = 2 \eta^{\mu \nu}$ with $\eta^{\mu \nu}$ mostly minus.  The chiral gamma matrix is defined to be $\gamma^5 = i \gamma^0 \gamma^1 \gamma^2 \gamma^3 $ and I work in the chiral basis where Dirac spinors decompose as $\psi^T = \left( \begin{array}{cc} \psi_L & \psi_R \end{array} \right)^T$ with
\begin{equation}
\gamma^5 = \left( \begin{array}{cc} - 1_2 & 0 \\ 0 & 1_2 \end{array} \right).
\end{equation}
The Weyl equation for a left handed spinor is
\begin{equation}
(i \partial_t - i \sigma^i D_i) \psi_L = 0,
\end{equation}
with $\sigma^i$ the usual Pauli matrices. The vector potential in Landau gauge is $A_y = B x $ for a constant magnetic field $B$ in the $z$ direction.  Most solutions of the Weyl equation in a finite magnetic field have a gap coming from the cyclotron motion, but there are also zero mode solutions.  These solutions may be heuristically understood as arising from a balance between the Zeeman energy and the orbital cyclotron energy.

Zero mode solutions may be found by putting $\partial_t \psi_L = \partial_z \psi_L = 0$ to obtain
\begin{equation}
\sigma^x \partial_x \psi_L + \sigma^y (\partial_y - i q B x ) \psi_L = 0.
\end{equation}
Landau gauge maintains translation invariance in the $y$ direction, so I try a solution of the form $\psi_L(x,y) = \psi_L(x) e^{i k y}$.  The Weyl equation reduces to
\begin{equation}
\partial_x \psi_L = - \sigma^z ( q B x - k ) \psi_L
\end{equation}
with solution
\begin{equation}
\psi_L(x) = \exp{\left(- \frac{qB}{2} \left(x - \frac{k}{qB}\right)^2 \sigma^z \right)} \psi_L(0).
\end{equation}
In order for this solution to be normalizable we must have $\sigma^z \psi_L (0) = \psi_L(0) $ (assuming $qB > 0$) leaving only one degree of freedom.  The spacing of $k$ is determined by the length of the system in the $y$ direction to be $\Delta k = \frac{2 \pi}{L_y}$.  We have one zero for each value of $k$ such that $\psi_L(x)$ sits inside the system in the $x$ direction.  The degeneracy $g$ of zero modes is thus $g = \frac{qB L_x}{\Delta k} = \frac{q B L_x L_y}{2\pi}$.  More generally, these zero modes and their degeneracy are protected by an index theorem relating the number of zero modes to the magnetic flux penetrating the system: $N_{\text{zero modes}} = \frac{q}{2 \pi} \int F_{12} dx dy$.

So far we have ignored the $z$ direction, but these zero modes actually disperse in the $z$ direction.  Assuming a more general solution of the form $\psi_L(x,y,z,t) = e^{i p_z z + i p_y y - i E t} \psi_L(x)$ the full Weyl equation becomes
\begin{equation}
E \psi_L - p_z \sigma^z \psi_L - i( \sigma^x \partial_x  + \sigma^y (\partial_y - i q B x )) \psi_L = 0.
\end{equation}
The second half of this equation is solved with the same zero mode profile as above.  The first half reduces to the equation $E = p_z$ using the fact that $\sigma^z \psi_L = \psi_L$ following from the normalization condition.  Thus each zero mode is actually relativistic chiral fermion in one spatial dimension.  The low energy physics is controlled entirely by these zero modes as all other modes are gapped by the cyclotron motion.

Using the one dimensional structure I can compute the entanglement entropy of the Weyl fermion.  Consider a box of linear size $L$.  The entanglement entropy $S_L$ is defined as the von Neumann entropy of the reduced density matrix corresponding to the box: $S_L = - \text{Tr}(\rho_L \ln{\rho_L})$. For one dimensional conformal field theories the entanglement entropy is known to have the form
\begin{equation}
S_L = \frac{c_L + c_R}{6} \ln{\left(\frac{L}{\epsilon}\right)},
\end{equation}
where $c_L$ and $c_R$ are the left and right central charges and $\epsilon$ is an ultraviolet cutoff \cite{eeqft}.  Weyl fermions in a magnetic field may be described by a large number of one dimensional gapless modes, and these modes are each equivalent to a chiral $1+1$ dimensional conformal field theory, the dimensions being $z$ and $t$.  Each chiral fermion mode has $c_L = 1$ and $c_R = 0$ and hence contributes $(1/6) \ln{L}$ to the entanglement entropy.  For a cube of side length $L$ aligned with the $z$ direction we have $q B L^2 / (2 \pi) $ zero modes for a total entanglement entropy
\begin{equation}
S_L = \left(\frac{q B L^2}{2 \pi} \right) \frac{1}{6} \ln{\left(\frac{L}{\epsilon}\right)}.
\end{equation}
This formula may be checked using the generalization of one dimensional entanglement entropy to finite temperature
\begin{equation}
S_L = \frac{c_L + c_R}{6} \ln{\left(\frac{\beta}{\pi \epsilon} \sinh{\frac{\pi L}{\beta} }\right)}.
\end{equation}
The thermal entropy of these zero modes in a cube of size $L$ is thus
\begin{equation}
S = \left(\frac{q B L^2}{2 \pi} \right) \frac{\pi L T}{6}
\end{equation}
which agrees with the direct thermodynamic calculation.

Before moving on, let me note that a single charged Weyl fermion does not give a consistent quantum theory.  This is due to the presence of a gauge anomaly proportional to $\text{Tr}(Q^3)$ where $Q$ is the charge matrix.  There is also a gravitational anomaly proportional to $\text{Tr}(Q)$.  Both of these anomalies must vanish for a completely well defined chiral gauge theory, but this can be accomplished by adding Weyl fermions with compensating charges.  The boundary law violating behavior remains, and thus there are consistent configurations of Weyl fermions that violate the boundary law for entanglement entropy.

\textit{Holographic generalization}.---I have computed the entanglement entropy for a single free Weyl fermion and found a term that violates the boundary law for entanglement entropy.  A useful choice for incorporating interactions is $\mathcal{N} = 4$ SU($N$) Yang-Mills theory which includes $4 N^2 $ Weyl fermions as part of the field content.  These fermions sit in the adjoint of the non-Abelian gauge group SU($N$), while the magnetic field $B$ corresponds to a weakly gauged U(1) subgroup of the R-symmetry.  In zero magnetic field this theory is conformal at all values of the t'Hooft coupling $\lambda = g^2_{YM} N$, but it is particularly amenable to study at strong coupling because of holographic duality.  This duality relates the $\mathcal{N} = 4$ theory to a theory of quantum gravity, IIB string theory, in an asymptotically five dimensional anti-de-Sitter spacetime ($\text{AdS}_5$).  The limit $\lambda \rightarrow \infty$ and $N \rightarrow \infty$ in the field theory gives classical supergravity in anti-de-Sitter space on the gravity side.

In this strong coupling limit, configurations of the super Yang-Mills theory have an emergent geometric interpretation in terms of classical gravitational field configurations.  The ground state of the field theory is dual to pure anti-de-Sitter space, and the field theory at finite temperature is accessed via a bulk black hole.  The field theory in a background magnetic field at zero temperature is obtained from a magnetically charged extremal black hole in the bulk.  Given the bulk geometric configuration, the leading large N contribution to the entanglement entropy can be determined holographically by computing the area of certain minimal surfaces in the bulk \cite{holo_ee,holo_ee_f}.

Consider extremal magnetic brane solutions in Einstein-Maxwell theory with negative cosmological constant in five dimensions \cite{mag_branes}.  These solutions interpolate between an asymptotically $\text{AdS}_5$ region and a near horizon $\text{AdS}_3 \times \text{T}^2$ region (assuming the xy plane is compactified).  The asymptotic $\text{AdS}_5$ region corresponds to the unperturbed $\mathcal{N} = 4$ theory at high energies.  The near horizon region appears as a result of turning on a magnetic field in the gauge theory.  The radial evolotion represents a renormalization group flow from a $3+1$ dimensional conformal field theory at high energies to an effectively $1+1$ dimensional conformal field theory at low energies.  This is qualitatively similar to the physics of free Weyl fermions, and even at strong coupling the cross-over scale is determined by the magnetic length.  At zero temperature the metric may be written in the form
\begin{equation}
ds^2 = - U(r)\, dt^2 + \frac{dr^2}{U(r)} + U(r)\, dz^2 + e^{2V}(dx^2 + dy^2),
\end{equation}
with $r$ the radial coordinate ($r \rightarrow \infty$ is the boundary) and $z$ the direction of the magnetic field on the boundary \cite{mag_branes}.  I use bulk units with the AdS radius set to one.  In addition to the metric, the gauge field has a profile given by $F = B\, dx \wedge dy $.  The asymptotic $\text{AdS}_5$ region is described by $U = e^{2V} = r^2$ while the near horizon $\text{AdS}_3 \times \text{T}^2$ region corresponds to $U = 3r^2$ and $e^{2V} = B/3$.  Notice that in the near horizon region the xy plane has decoupled from the radial coordinate and has fixed size given by the magnetic length.

The entanglement entropy of a region in the dual field theory is determined by the area of the minimal surface in the bulk that terminates on the boundary of the region in the field theory.  The entanglement entropy is just this minimal area divided by $4 G^{(5)}_N$.  I will focus on the entanglement entropy of a rectangular region in boundary theory of size $L \times L \times L_z$.  Assuming $L \gg L_z$ gives approximate translation invariance in the xy plane.  The minimal surface calculation reduces to a two dimensional problem involving only the variables $z$ and $r$.  The zero temperature geometry is only known numerically, and the minimal surface calculation can also only be done numerically.  However, the important physics can be extracted without the numerical details.  For cubic regions with all dimensions less than the magnetic length, the minimal surface only probes the $\text{AdS}_5$ region and gives the usual ultraviolet divergent boundary law for entanglement entropy.

For boundary regions of linear size much larger than the magnetic length, the minimal surface passes right through the asymptotic $\text{AdS}_5$ region towards the near horizon region.  Once in the near horizon region, the $x$ and $y$ directions freeze out, and the minimal surface behaves exactly as in AdS$_3$.  In particular, I find the characteristic $\ln{(L_z / \ell)}$ dependence familiar from $1+1$ dimensional conformal field theory with the magnetic length providing the cutoff.  The entanglement entropy thus consists of two pieces, a non-universal boundary law contribution from the asymptotically $\text{AdS}_5$ region and a universal low energy piece $S_L \sim N^2 B L^2 \ln{(L_z / \ell_B)}$.  The appearance of the magnetic field can be understood because the effective $1+1$ dimensional central charge is related to $1/G^{(3)}_N$ which is enhanced relative to $1/G^{(5)}_N$ by a factor of $B L^2$ from the freeze out of the xy plane.  This strong coupling version of the free Weyl fermion system thus also violates the boundary law for entanglement entropy at low energies.

\textit{Strong topological insulators}.---In both the cases considered above, the appearance of many gapless one dimensional modes was responsible for the highly entangled nature of the quantum state.   This intuition can be applied to more experimentally relevant systems known as strong topological insulators.  These systems are time reversal invariant electronic band insulators that are not smoothly connected to trivial band insulators.  In particular, they possess interesting topological structure that gives rise to protected edge modes.  These edge modes are robust so long as time reversal invariance is preserved \cite{ti1,ti4,ti10}.  Topological insulators in three spatial dimensions have gapless surface states living in two spatial dimensions, but these modes do not lead to a violation of the boundary law for entanglement entropy.  Similarly, the bulk of a topological insulator is gapped in a perfect crystal and certainly satisfies a boundary law for entanglement entropy.

However, experimentally realized topological insulators are not perfect crystals, they possess topological defects including dislocations in the crystalline bulk. Remarkably, for certain kinds of topological insulators and dislocation types, the dislocations have been shown to support gapless fermionic modes \cite{helical_metal}.  These effectively one dimensional modes make the dislocations into gapless quantum wires threading the otherwise gapped bulk.  The one dimensional modes in the quantum wires are analogous to the Weyl zero modes considered above, with the dislocations playing the role of magnetic field lines.  In the presence of a finite density of dislocations supporting gapless modes, the bulk of a strong topological insulator violates the boundary law for entanglement entropy.

To estimate the size of the violation, consider the artificial situation of a dilute array of topologically non-trivial dislocations all aligned.  Let these dislocations have an areal density $\rho$ (a typical value of $\rho$ might be $10^{12}\, \text{m}^{-2}$  \cite{helical_metal}).  A region in the bulk of size $L \times L \times L_z$, with the z axis chosen parallel to the dislocations, effectively contains $\rho L^2 $ gapless one dimensional fermionic modes.  These modes should each contribute roughly $\ln{(L_z / \epsilon)}$ to the entanglement entropy.  The boundary law violating component of the entanglement entropy is thus of order $S_L \sim \rho L^2 \ln{(L_z / \epsilon)}$.  This estimate is crude, but it should suffice for a reasonably uniform and collimated set of dislocations.  Note that despite the enhanced $L$ dependence relative to the usual boundary law, this term may be much smaller than the boundary law term for experimentally accessible system sizes and dislocation densities.  I also wish to emphasize that this is a statement about the zero temperature quantum state.  The helical modes modes are protected from elastic scattering (such scattering might otherwise localize a one dimensional gapless mode), but at finite temperature or in the presence of inelastic processes, the boundary law violating behavior will be disrupted.

\textit{Discussion}.---I have described three systems in $3+1$ dimensions that violate the boundary law for entanglement entropy.  The mechanism for these violations is the emergence of a low energy description in terms of many gapless one dimensional modes.  A completely analogous picture has been argued to hold for Fermi liquids in any dimension, with the exact coefficient of the boundary law violating term determined by the geometry of the interacting Fermi surface.  The generic picture of modes propagating along one dimensional objects, be they magnetic field lines or topological defects, permits a unified understanding of the weakly coupled systems I considered.  In the strong coupling limit, holographic duality provides a geometric realization of a similar emergent $1+1$ dimensional description in the magnetic field direction and time.

These systems are interesting in part because they expand our understanding of the limits of the boundary law for entanglement entropy.  Currently, violations of the boundary law in reasonable (without large ground state degeneracy) translation invariant systems can be traced to a one dimensional picture.  The one dimensional physics in these systems suggests the possibility of enhanced fluctuations associated with violations of the boundary law \cite{fermion2,emeter,fs_fluc_1d,bgs_ferm2}.  A different test might come from thermodynamic measurements.  The thermal signature of boundary law violating entanglement entropy in these effectively relativistic one dimensional systems is a contribution to heat capacity linear in $T$.  In the case of topological insulators, this term will be quite small, but it does have a different temperature dependence than the contributions from surface states, gapped electrons, and phonons.

There are many directions for future work.  In the direction of classification, it is an interesting open question whether all violations of the boundary law in reasonable translation invariant systems can be traced to one dimensional physics.  Are there examples of reasonable quantum systems even more highly entangled than these effectively one dimensional systems?  On the experimental front, it has proved difficult to find definite signatures of entanglement in many body systems in higher dimensions.  The situation is better in one dimension where the entanglement entropy can be directly measured using current fluctuations \cite{emeter}.  It would be interesting to explore in detail the experimental consequences of entanglement in higher dimensional systems that are nevertheless controlled by one dimensional physics.

\bibliography{weyl_ee}

\end{document}